\documentstyle[prd,aps,preprint,epsfig]{revtex}

\tighten

\begin{document}
\draft
 
\pagestyle{empty}

\preprint{
\noindent
\hfill
\begin{minipage}[t]{3in}
\begin{flushright}
August 1999
\end{flushright}
\end{minipage}
}

\title{Testing the direct CP violation of 
the Standard Model without knowing strong phases}

\author{
Mahiko Suzuki
}
\address{
Department of Physics and Lawrence Berkeley National Laboratory\\
University of California, Berkeley, California 94720
}


\maketitle

\begin{abstract}

The strong phase of a two-body decay amplitude of a heavy particle
depends on decay operators even if the final state is an eigenstate 
of isospin or SU(3) quantum numbers. This particular property 
extends the opportunity of testing consistency of 
experimentally observed {\em CP}-violation phases with
the Standard Model without knowing strong interaction effects in
decay amplitudes. With three generations, the Standard Model requires
$\Delta(\pi^{\pm}\eta')=-\Delta(K^{\pm}\eta')$ in the flavor SU(3) 
symmetry limit, where $\Delta(\pi^{\pm}\eta')\equiv 
{\rm B}(B^+\rightarrow\pi^+\eta')-{\rm B}(B^-\rightarrow\pi^-\eta')$ 
and $\Delta(K^{\pm}\eta')\equiv{\rm B}(B^+\rightarrow K^+\eta')-
{\rm B}(B^-\rightarrow K^-\eta')$. However, testing this relation with
the Standard Model is not easy. The relation $\Delta(\pi^{\pm}\psi)
= -\Delta(K^{\pm}\psi)$ is cleaner but even harder to test.

\end{abstract}
\pagestyle{plain}
\narrowtext

\setcounter{footnote}{0}

\section{Introduction}

In the minimal Standard Model with three generations (MSM) there is only
one independent {\em CP}-violating parameter.  Therefore, in principle,
determining the weak phase $\beta$ from the {\em CP} violating 
$B^0-\overline{B}^0$ mixing is sufficient within the MSM. One of the major
purposes of exploring for the phase $\gamma$ in direct {\em CP} violation 
processes is to test consistency of other {\em CP} violation phenomena 
with the MSM and to search for possible sources of {\em CP} violation 
beyond the MSM. 

Many proposals have been made as to how to extract the phase $\gamma$ 
from direct {\em CP}-violating processes\cite{GR}. The difficulty
is that the weak phases are entangled with unknown strong phases due to
final state interactions (FSI). In many cases, 
one can in principle determine both weak and strong phases by measuring
sufficiently many decay modes.  Since experimental errors accumulate 
with the number of measured values, however, an unrealistically 
high precision is often required for measurement. Use of flavor SU(3) 
symmetry is a powerful way to simplify the theoretical analysis 
by reducing the number of independent decay amplitudes.  Nevertheless,
additional dynamical approximations and/or assumptions are needed
to make the extraction of $\gamma$ feasible. While a model
such as the factorization model may give us some idea of the relative 
magnitudes of decay amplitudes, the strong phases of amplitudes are
much harder to compute unless short-distance QCD should
completely dominate.\footnote{If the strong phases of two-body $B$ 
decay are dominated by short-distance QCD\cite{BJ,Beneke}, 
all strong phases would be small and calculable in principle. However, 
a convincing {\em quantitative} proof is yet to be given for the 
short-distance dominance.}  Because of the uncertainty of strong 
phases some are content only with setting bounds on the phase $\gamma$.
 
In order to extract the weak phases from direct {\em CP} violations, we 
need a set of decay modes which are described by two or more of 
independent decay amplitudes differing in the strong phase. The fewer the 
independent amplitudes are, the simpler the analysis is. We would like 
to avoid theoretical assumptions and approximations on those decay 
amplitudes as much as possible, preferably treating them as free 
parameters without theoretical prejudice. For this reason, we should 
study a set of decay modes that involves the smallest number 
of independent amplitudes. With SU(3) symmetry He recently derived
several relations\cite{He} for the rate differences of the two-body octet
pseudoscalar-meson decay modes which do not depend on strong interaction
effects at all. The final states considered by He contain two or more 
of isospin or SU(3) eigenstates to generate a strong phase difference. 
However, high inelasticity and multichannel coupling of the final 
states of the {\em B} decay make a {\em CP} asymmetry observable even 
in the final states which are eigenstates of isospin or SU(3). 
We shall briefly remind this important fact in Section II in
order to add a few more promising relations of the same nature to
the list of \cite{He}. In Section III, we derive the relation for the rate 
differences of singlet-octet two-body final states, which are not only 
isospin eigenstates but also octet eigenstates of SU(3).  
Comments will be made on feasibility of test in Section IV.

\section{Final state interaction}

    When many decay channels are open in a heavy particle decay, 
the FSI phases of decay amplitudes for experimentally 
measured final-states are not simply related to the phases of
pure strong interaction. Take, for example, a two-body final state 
$|ab\rangle$. The state $|ab\rangle$ is not one of the eigenstates
$|\alpha\rangle$ of strong interaction {\em S} matrix.
When the eigenchannels of {\em S} matrix are defined by
\begin{eqnarray}
    \langle\beta|S|\alpha\rangle
         &=& \langle\beta^{{\rm out}}|\alpha^{{\rm in}}\rangle ,\\ \nonumber 
         &=& \delta_{\beta\alpha}e^{2i\delta_{\alpha}},
\end{eqnarray} 
an experimentally observable final-state is a linear combination 
of them. Take, for example, a two-body final state $|ab\rangle$. 
The state $|ab\rangle$ is expanded as
\begin{equation}
       |ab\rangle = \sum_{\alpha}O_{ab,\alpha}|\alpha\rangle. \label{O}
\end{equation} 
Time reversal invariance of strong interaction allows us to choose the
{\em S} matrix to be symmetric and $O_{ab,\alpha}$ to be an orthogonal matrix.
For a {\em CP}-even decay operator $\cal O$$_i$, time-reversal operation
leads us to
\begin{equation}
     \langle\alpha^{{\rm out}}|{\cal O}_i|B\rangle =
            \langle B|{\cal O}_i|\alpha^{{\rm out}}\rangle
            \langle\alpha^{{\rm out}}|\alpha^{{\rm in}}\rangle.
\end{equation}
Therefore the decay amplitude takes the form
\begin{equation}
     \langle\alpha^{{\rm out}}|{\cal O}_i|B\rangle = 
                             M_{\alpha}^ie^{i\delta_{\alpha}},
\end{equation}
where
\begin{equation}
       (M_{\alpha}^i)^* = M_{\alpha}^i.
\end{equation}
This is the well-known phase theorem\cite{Watson} in the case that
the final state is an eigenstate of {\em S} matrix.  When $|ab\rangle$ 
is not an eigenstate of {\em S} matrix, but is given by Eq. (\ref{O}), 
the decay amplitude for $B\rightarrow ab$ is a superposition of 
$B\rightarrow\alpha$:
\begin{equation}
  M^i(B\rightarrow ab)=
         \sum_{\alpha}O_{ab,\alpha}e^{i\delta_{\alpha}}M_{\alpha}^i.
                    \label{FSI}
\end{equation}

We should learn two important facts from Eq. (\ref{FSI}). One is that 
the net (strong) phase of $M(B\rightarrow ab)$ is not simply related 
to the eigenphase shifts $\delta_{\alpha}$ of {\em S} matrix. It is not 
given by a phase of any pure strong interaction process, elastic 
or inelastic, of $|ab\rangle$. The other is that the phase of 
$M(B\rightarrow ab)$ is dependent on the operator $\cal O$$_i$. 
For instance, the strong phase of the $B\rightarrow K\pi$ 
amplitude into total isospin 1/2 takes different values for the tree 
decay process and for the penguin decay process. There is no reason to 
expect that the two values are even close to each other, since the different 
quark structures of $\cal O$$_{1,2}$ and $\cal O$$_{3\sim 10}$ generate
very different sets of $M_{\alpha}^i$ in general. The strong phases
of the tree and the penguin amplitude of $(K\pi)_{I=1/2}$
can be just as much different as those of $(K\pi)_{I=1/2}$ and 
$(K\pi)_{I=3/2}$ are, or as those of $(K\pi)_{{\bf 8}}$
and $(K\pi)_{{\bf 27}}$ of SU(3) are. 

Thanks to this property of the FSI in the {\em B} decay, the {\em CP} asymmetry
can appear even in an isospin eigenstate or an SU(3) eigenstate.
A merit of considering such final states is that since their strong 
interaction parametrization is very simple, we can more easily disentangle 
the weak phases from the strong phases.

\section{SU(3) analysis}

We cast the effective Hamiltonian of the {\em B} decay into the form 
\begin{equation}
     H_{eff}\simeq 2\sqrt{2}G_F\sum_{q=d,s}
               (V_{ub}V_{uq}^* \sum_{i=1}^{2}C_i{\cal O}_i^q
      -V_{tb}V_{tq}^* \sum_{j=3}^{10}C_j{\cal O}_j^q) + {\rm H.c.},
\end{equation}
where the decay operators are defined by
\begin{eqnarray}
 {\cal O}_1^q &=&(\overline{u}\gamma^{\mu}b_L)
                      (\overline{q}\gamma_{\mu}u_L)
                  - (\overline{c}\gamma^{\mu}b_L)
                      (\overline{q}\gamma^{\mu}c_L), \label{O1} \\  
 {\cal O}_2^q &=&(\overline{q}\gamma^{\mu}b_L)
                      (\overline{u}\gamma_{\mu}u_L)
                  - (\overline{q}\gamma^{\mu}b_L)
                      (\overline{c}\gamma_{\mu}c_L) , \\ 
 {\cal O}_3^q &=&\sum_{q'=u,d,s,c}(\overline{q}\gamma^{\mu}b_L)
                      (\overline{q'}\gamma_{\mu}q'_L)
                      +\frac{C_2}{C_3}(\overline{q}\gamma^{\mu}b_L)
                      (\overline{c}\gamma_{\mu}c_L), \\ 
{\cal O}_4^q &=&\sum_{q'=u,d,s,c}(\overline{q_{\alpha}}\gamma^{\mu}
           b_{\beta L})(\overline{q'_{\beta}}\gamma_{\mu}q'_{\alpha L})
             +\frac{C_1}{C_4}(\overline{q_{\alpha}}\gamma^{\mu}b_{\beta L})
            (\overline{c_{\beta}}\gamma_{\mu}c_{\alpha L}),  \label{O4}\\  
 {\cal O}_5^q &=&\sum_{q'=u,d,s,c}(\overline{q}\gamma^{\mu}b_L)
                      (\overline{q'}\gamma_{\mu}q'_R), \\
{\cal O}_6^q &=&\sum_{q'=u,d,s,c}(\overline{q_{\alpha}}\gamma^{\mu}
           b_{\beta L})(\overline{q'_{\beta}}\gamma_{\mu}q'_{\alpha R}), \\
 {\cal O}_7^q &=&\frac{3}{2}\sum_{q'=u,d,s,c}
                      (\overline{q}\gamma^{\mu}b_L)
                   e_{q'}(\overline{q'}\gamma_{\mu}q'_R), \\
 {\cal O}_8^q &=&\frac{3}{2}\sum_{q'=u,d,s,c}
                      (\overline{q_{\alpha}}\gamma^{\mu}b_{\beta L})
                   e_{q'}(\overline{q'_{\beta}}\gamma_{\mu}q'_{\alpha R}), \\
 {\cal O}_9^q &=&\frac{3}{2}\sum_{q'=u,d,s,c}
                      (\overline{q}\gamma^{\mu}b_L)
                   e_{q'}(\overline{q'}\gamma_{\mu}q'_L), \\
 {\cal O}_{10}^q &=&\frac{3}{2}\sum_{q'=u,d,s,c}
                      (\overline{q_{\alpha}}\gamma^{\mu}b_{\beta L})
                   e_{q'}(\overline{q'_{\beta}}\gamma_{\mu}q'_{\alpha L}).
\end{eqnarray}
In grouping the terms in $H_{eff}$, we have expressed the 
coefficient $V_{cb}V_{cq}^*$ of the tree operators involving $c$ and 
$\overline{c}$ in terms of $V_{ub}V_{uq}^*$ and $V_{tb}V_{tq}^*$ by
using the unitarity relations of three generations,
\begin{equation}
     V_{ub} V_{uq}^* + V_{cb} V_{cq}^* + V_{tb} V_{tq}^* = 0,\;\;(q=d,s),
\end{equation}
and have distributed them into $\cal O$$_{1\sim 4}$ in Eq. (\ref{O1})
$\sim$ (\ref{O4}). The tree operators of $c\overline{c}$ are 
potentially important if the FSI should allow a substantial conversion 
of $c\overline{c}\rightarrow$ light quark pairs\cite{Ciuchini}.

It is important to notice here that all decay operators 
(${\cal O}$$_i^d$, ${\cal O}$$_i^s$) ($i=1\sim 10$) form
doublets under the {\em U}-spin rotation ($d\leftrightarrow s$) of an 
SU(3) subgroup. Under {\em U}-spin, $B^{\pm}$ are singlets while 
$(B^0, B_s^0)$ forms a doublet. Likewise $(\pi^-, K^-)$ is a 
doublet.\footnote{This 
{\em U}-spin property immediately leads to six  
of the relations written in \cite{He}}. Here we consider the $B^{\pm}$ 
decay into $\pi^{\pm}\eta'$ and $K^{\pm}\eta'$ instead of the 
$B^0/\overline{B}^0$ and $B_s/\overline{B}_s$ decays: 
\begin{eqnarray}
      B^{\pm} &\rightarrow& K^{\pm}\eta',   \\      \label{mode1} 
      B^{\pm} &\rightarrow& \pi^{\pm}\eta'.        \label{mode2}
\end{eqnarray}
In the SU(3) symmetry limit leaving out the $\eta-\eta'$ mixing,
the decay amplitudes for $B^{\pm}\rightarrow\pi^{\pm}\eta'$ and 
$K^{\pm}\eta'$ are parametrized in the form
\begin{eqnarray}
  M(\pi^+\eta') &=& V_{ud}V_{ub}^*T + V_{td}V_{tb}^*P,   \label{amp1}\\  
  M(K^+\eta')   &=& V_{us}V_{ub}^*T + V_{ts}V_{tb}^*P,   \label{amp2}
\end{eqnarray}
where 
\begin{eqnarray}
 T&=&2\sqrt{2}G_F\langle \pi^+\eta'|\sum_{i=1}^2 
                          C_i{\cal O}_i^{\dagger}|B^+\rangle \\
 P&=&2\sqrt{2}G_F\langle \pi^+\eta'|\sum_{j=3}^{10}
                          C_j{\cal O}_j^{\dagger}|B^+\rangle.
\end{eqnarray}
The QCD and electroweak penguin contributions have been combined 
into a single term
\begin{equation}
       P = P_{QCD}+P_{EW}.
\end{equation}
The decay amplitudes for $B^-\rightarrow\pi^-\eta'$ and $K^-\eta'$
are obtained from Eqs. (\ref{amp1}) and (\ref{amp2}) by complex
conjugation of the quark mixing matrix elements.

The FSI turns the amplitudes $T$ and $B$ complex and, according to 
our argument in Section II, their phases are different from each other
in general. Therefore the rate differences
\begin{eqnarray}
     \Delta(\pi^{\pm}\eta'(K^{\pm}\eta'))&=&
   {\em B}(B^+\rightarrow\pi^+\eta'(K^+\eta'))-
         {\em B}(B^-\rightarrow\pi^-\eta'(K^-\eta'))  \\
                &=& 4|T||P|\sin\delta\theta\; 
         {\rm Im}(V_{uq}V_{ub}^*V_{tb}V_{tq}^*) \;\; (q= d,s),
\end{eqnarray} 
where $\delta\theta =\arg(T^*P)$, are nonvanishing. Though the final 
states are isospin eigenstates, $\Delta(\pi^{\pm}\eta')$ and 
$\Delta(K^{\pm}\eta')$ can be just as large as those of isospin 
non-eigenstates. The imaginary part of the product of the
quark mixing matrix elements is common to $q=d$ and $s$ up to 
a sign\cite{Jarlskog}:
\begin{equation}
   {\rm Im}(V_{ud}V_{ub}^*V_{tb}V_{td}) = 
      - {\rm Im}(V_{us}V_{ub}^*V_{tb}V_{ts}).
\end{equation}
We thus come to the relation,
\begin{equation}
     \Delta(\pi^{\pm}\eta') = - \Delta(K^{\pm}\eta'). \label{Delta}
\end{equation}
This relation is not useful in extracting the weak phase $\gamma$ unless
we know $|T||P|$ and $\delta\theta$ beforehand from somewhere else.
From the viewpoint of testing {\em CP} violations in the MSM, however, 
it is one of the cleaner tests and will serve the same goal as determining 
$\gamma$ through complex procedures.

\section{Comments on SU(3) breaking} 

The $K^{\pm}\eta'$ mode is the largest in branching fraction among
all charmless two-body $B^{\pm}$ decay modes so far measured. The
$\pi^{\pm}\eta'$ mode has not been measured. In a theoretical analysis
based on SU(3)\cite{Dighe}, $\pi^{\pm}\eta'$ is expected to be
competitive with $\pi^{\pm}\pi^0$ and to be one of the largest in
branching fraction among the flavorless final states. Measurability
of a {\em CP} asymmetry in $\pi^{\pm}\eta'$ was actually pointed out
by the authors of \cite{Dighe} and \cite{Barshay}. The competitive rates
of $K^{\pm}\eta'$ and $\pi^{\pm}\eta'$ may give an advantage to
Eq. (\ref{Delta}) over the relation 
$\Delta(\pi^+K^0/\pi^-\overline{K}^0)=
-\Delta(K^+\overline{K}^0/K^-K^0)$ of \cite{He}.

We have ignored SU(3) breaking of strong interaction in Eq. (\ref{Delta}).
It is likely that the SU(3) breaking in rescattering dynamics is 
insignificant at the energy of {\em B} mass. In the factorization
model, the SU(3) breaking associated with each meson can be incorporated 
by $\Delta\rightarrow f_{\pi(K)}\Delta$. We shall learn more about
reliability of factorization by comparing the theoretical predictions 
with experiment\cite{Dighe}.  

     The $\eta-\eta'$ mixing is one manifestation of SU(3) breaking.  
This may be viewed as a disadvantage of our relation. Recently 
a dynamical model was proposed to compute the decay matrix elements
of $B^{\pm}\rightarrow \pi^{\pm}\eta'$ and $K^{\pm}\eta'$\cite{Almady}.  
In this model $eta'$ is generated through two gluons in the penguin diagrams
while $u\overline{u}$ forms $\eta'$ in the tree diagrams as a color-favored
process.  If these processes are the dominant ones, the $\eta-\eta'$ mixing 
correction appears as a common factor on both sides of Eq .(\ref{Delta})
and does not affect the relation.

  Since we expect ${\rm B}(B^{\pm}
\rightarrow K^{\pm}\eta')$ to be much larger than ${\rm B}(B^{\pm}
\rightarrow\pi^{\pm}\eta')$, a small difference between two large
numbers will be searched for in the right-hand side of Eq. (\ref{Delta}),
while the left-hand side will be obtained hopefully as a fairly large
difference between two smaller numbers.
If we take the estimates by the authors of \cite{Almady} as a ballpark
figure, their preferred values for $B^{\pm}\rightarrow\pi^{\pm}\eta'$ 
lead to $\Delta(\pi^+\eta')-\Delta(\pi^-\eta') \simeq 4\times 10^{-6}$
which corresponds to a 40\% asymmetry. 
Then we shall be looking for a 3\% of asymmetry in 
the $K^{\pm}\eta'$ mode up to a possible 22\% upward correction due 
to $f_K/f_{\pi}$.  If this is the case, testing the relation with the MSM 
will be rather a remote possibility in the B factory experiment. 

The same relation as Eq. (\ref{Delta}) should hold for
$B^{\pm}\rightarrow\rho^{\pm}\eta'$ and $K^{*\pm}\eta'$:
\begin{equation}
  \Delta(\rho^{\pm}\eta') = - \Delta(K^{*\pm}\eta').   
\end{equation} 
We can replace $\rho^{\pm}$ and $K^{*\pm}$ with the corresponding
components of any meson octet, respectively.

Finally, it is tempting to try for 
$B^{\pm}\rightarrow\pi^{\pm}\psi$ and $K^{\pm}\psi$
\begin{equation}
 \Delta(\pi^{\pm}\psi)=-\Delta(K^{\pm}\psi),
\end{equation}
since the relation is free from the $\eta-\eta'$ mixing 
contamination. Here again we may replace $\pi^{\pm}$ and $K^{\pm}$
with the corresponding components of any meson octet.
Furthermore, the rates are high and the 
experimental signature of $l^+l^-\pi^{\pm}(K^{\pm})$ 
is very clean. Unfortunately the asymmetries are will be even smaller. 

\acknowledgements
This work was supported in part by the Director, Office of Science, Office of
High Energy and Nuclear Physics, Division of High Energy Physics, of the
U.S. Department of Energy under Contract DE--AC03--76SF00098 and in
part by the National Science Foundation under Grant PHY--95--14797.



 
\end{document}